# The Core & Periphery Hypothesis:
# A Conceptual Basis for Generality in Cell and Developmental Biology

Elisa Gallo[1], Stefano De Renzis[2], James Sharpe[3,4],
Roberto Mayor[5], and Jonas Hartmann[1,2,3,5,6,7,*]

[1]Institute of Molecular Life Sciences, University of Zurich (UZH), 8057 Zurich, Switzerland

[2]Developmental Biology Unit, European Molecular Biology Laboratory (EMBL), 69117 Heidelberg, Germany

[3]EMBL Barcelona, European Molecular Biology Laboratory (EMBL), 08003 Barcelona, Spain

[4]Institució Catalana de Recerca i Estudis Avançats, 08010 Barcelona, Spain

[5]Department of Cell and Developmental Biology, University College London, WC1E 6BT London, UK

[6]Northwestern Institute on Complex Systems, Northwestern University, 60208 Evanston IL, USA

[7]Department of Chemical and Biological Engineering, Northwestern University, 60208 Evanston IL, USA

*Correspondence: Jonas Hartmann (jonas.hartmann@ucl.ac.uk)

## Abstract

The discovery of general principles underlying the complexity and diversity of cellular and developmental systems is a central and long-standing aim of biology. Whilst new technologies collect data at an ever-accelerating rate, there is growing concern that conceptual progress is not keeping pace. We contend that this is due to a paucity of appropriate conceptual frameworks to serve as a basis for general theories of mesoscale biological phenomena. In exploring this issue, we have developed a foundation for one such framework, termed the Core and Periphery (C&P) hypothesis, which reveals hidden generality across the diverse and complex behaviors exhibited by cells and tissues. Here, we present the C&P concept, provide examples of its applicability across multiple scales, argue its consistency with evolution, and discuss key implications and open questions. We propose that the C&P hypothesis could unlock new avenues of conceptual progress in cell and developmental biology.

## Introduction

Cell and developmental biology aim to describe, understand, predict, and control the phenomena taking place in living systems at the mesoscopic scales between molecules and organisms. Progress toward these aims is driven on one hand by the collection of empirical data and on the other hand by the formulation of models and principles to explain said data. In recent years, the rate of data collection has accelerated tremendously, driven by rapid technological development both within and outside of biology (*1–5*). However, it does not appear that the search for explanatory principles has kept pace with this trend. Indeed, there are growing concerns that the conceptual understanding of mesoscale biology has advanced comparatively little, a claim that is of course hotly debated (*6–16*).

Advances in conceptual understanding are dependent on overarching **conceptual frameworks** (see Glossary of Terms) that enable the interpretation and generalization of new results, facilitate formal and informal reasoning, and guide experimental research by aiding in the conception and selection of research questions and working hypotheses (*17–21*). In mesoscale biology, the most prominent such framework in the past three decades has been the **gene-function paradigm** (Fig. 1a). Shaped by the triumph of genetic screening (*22–24*), this view posits that mesoscale biological functions can be understood by mapping them to specific genes and vice versa. Importantly, this framework appeared both simple and general, and therefore seemed to provide a fruitful basis for the deconstruction of complicated biological phenomena and for the (gene conservation-based) generalization of insights across species.

With time, however, it became increasingly clear that this linear interpretation of the gene-function paradigm does not adequately capture the complex behaviors of cellular and multi-cellular systems (*25–29*). Pleiotropy and epistasis were found to be common in gene-function relationships (Fig. 1b), revealing that such mappings are far from simple and that the functions of conserved genes are often heavily context-dependent and thus do not generalize well (*30–33*). We surmise that this loss of simplicity and generality initiated a gradual shift in the perspective of cell and developmental biologists toward a layered interpretation (Fig. 1c) in which the term "gene function" pertains mainly to the molecular function of a gene product (e.g. signal transduction or transcriptional activation), whereas higher-level functions are better described by the collective behavior of many parts that together form a system, such as proteins forming a signaling network or cells forming a tissue.

Although this shift toward a systems perspective constitutes some form of conceptual progress, it has not been accompanied by the widespread adoption of systems-level conceptual frameworks that could serve as a basis for **general** explanations and theories of mesoscale biological phenomena (*7, 15, 34–37*). This is despite strong and eloquent advocacy for the pursuit of such **generality** even before the rise of developmental genetics (by e.g. Wolpert in (*38*)), and despite interesting conceptual work along those lines having been done since (see e.g. (*39–43*)). In consequence, systems-level explanations today are often constructed in an *ad hoc* fashion and end up being just as idiosyncratic to a specific context as the classical notion of gene function. This severely limits our ability to compare or extrapolate explanations across biological systems, and by extension our ability to control or design them.

Exploring this issue with a focus on the complex behavioral repertoire of cells and cell collectives led us to develop the Core & Periphery (C&P) hypothesis, which we propose to be a suitable candidate for a novel conceptual framework in mesoscale biology. Here, we introduce the C&P concept and describe how it applies to biological systems from the subcellular to the embryonic scale. We then briefly cover evolutionary considerations that support the hypothesis and its ability to serve as a basis for theories that generalize across varied biological systems. Finally, we discuss a number of predictions, implications and open questions that follow from taking a C&P perspective. While the C&P framework remains a hypothesis until its claims have been tested empirically, we believe that its further study and application could accelerate conceptual progress in cell and developmental biology.

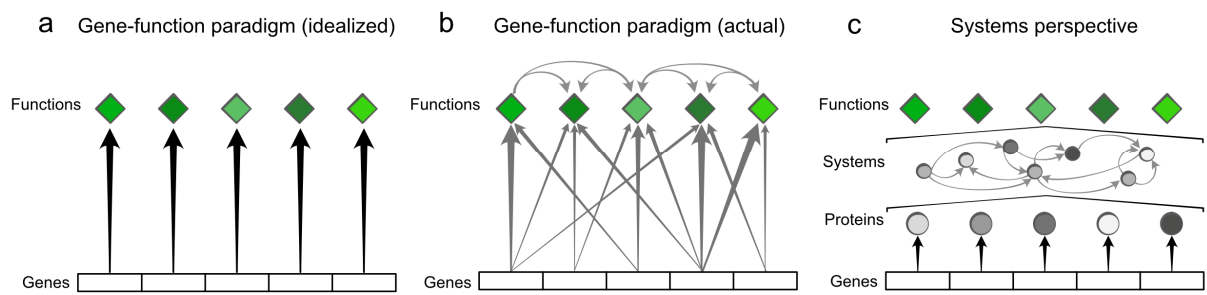

**Figure 1: Perspectives on cell and developmental biology**

**(a)** The gene-function paradigm in its simple and general form seeks to assign functions to genes and vice versa. **(b)** Despite the success of the gene-function paradigm, it has become increasingly clear that mesoscale biological functions usually involve many genes, and that most genes contribute to many such functions. In addition, many higher-order biological phenomena are better described using concepts other than genes (*curved gray arrows*), such as network motifs, cell behaviors, or physical forces. Gene function above the molecular scale is thus highly context-dependent. **(c)** As a consequence, many biologists view functions above the molecular scale as mediated by complex systems (*middle layer*). However, our understanding of such systems still tends to be context-specific, as there is a dearth of widely-applicable conceptual frameworks that could serve as a basis for generality in systems-level biological explanations.

# Main

### The Core & Periphery (C&P) hypothesis

Cell and developmental biology encompass a broad collection of qualitatively different phenomena that occur in myriad variations. Finding conceptual frameworks that generalize across such vast diversity seems near-impossible. However, in Darwin's time the same could have easily been said at the scale of organisms and populations, yet evolutionary theory succeeded in accounting for much of their complexity and diversity through a relatively simple process that can bootstrap itself from basic error-prone reproducers. Importantly for our purposes, evolutionary theory does not achieve this by stripping away the particulars of different species to find some lowest common denominator, but rather by formulating a **generative principle** that explains how the diversity of said particulars comes to be. This led us to conjecture that generative principles should likewise play a key role in conceptual frameworks for mesoscale biology. Indeed, it has previously been suggested in broader terms that generative systems tend to become deeply entrenched (and thus general) in evolutionary processes, including in animal development and in cultural evolution (*44*, *45*).

With this in mind, it is intriguing to note that certain kinds of systems are intrinsically capable of generating a wide range of different behaviors or outputs. So-called Turing systems are a well-known biological example, wherein the differential diffusion of positive and negative feedback signals gives rise to an instability that can generate qualitatively different patterns (spots, stripes, labyrinths) in endless quantitative variations (*46–49*). Outside of biology, modern computers are another (though very different) example. Whilst ordinary electronic circuits are engineered for only one specific purpose, computers are based on general-purpose circuits that can compute many different functions. Such systems with high inherent **versatility** can be directed to produce one particular pattern or function by an encompassing periphery that provides instructions or constraints, supplies appropriate input, and interprets the output. In computing, this periphery is mediated by software programs and by connected devices, such as keyboards and printers. In Turing systems in embryonic development, it is mediated by initial conditions such as upstream signals, by boundary conditions such as tissue geometry, and by the cells that read out and respond to the resulting morphogen patterns (*50–53*).

We refer to systems with this architecture – consisting of an inherently versatile system core embedded in a function-specific system periphery that "programs" it – as **Core & Periphery systems** (Fig. 2a-c). To be exact, we define a **system core** to be a subset of a biological system that has the intrinsic capacity to generate a wide range of non-trivial behaviors. Conversely, we define a **system periphery** to be the subset of a biological system that is not part of the core and instead triggers or programs it to perform one specific functional behavior out of the many that it potentially could. We expect cores to have a highly non-linear and integrated structure (such as the tight feedback within a

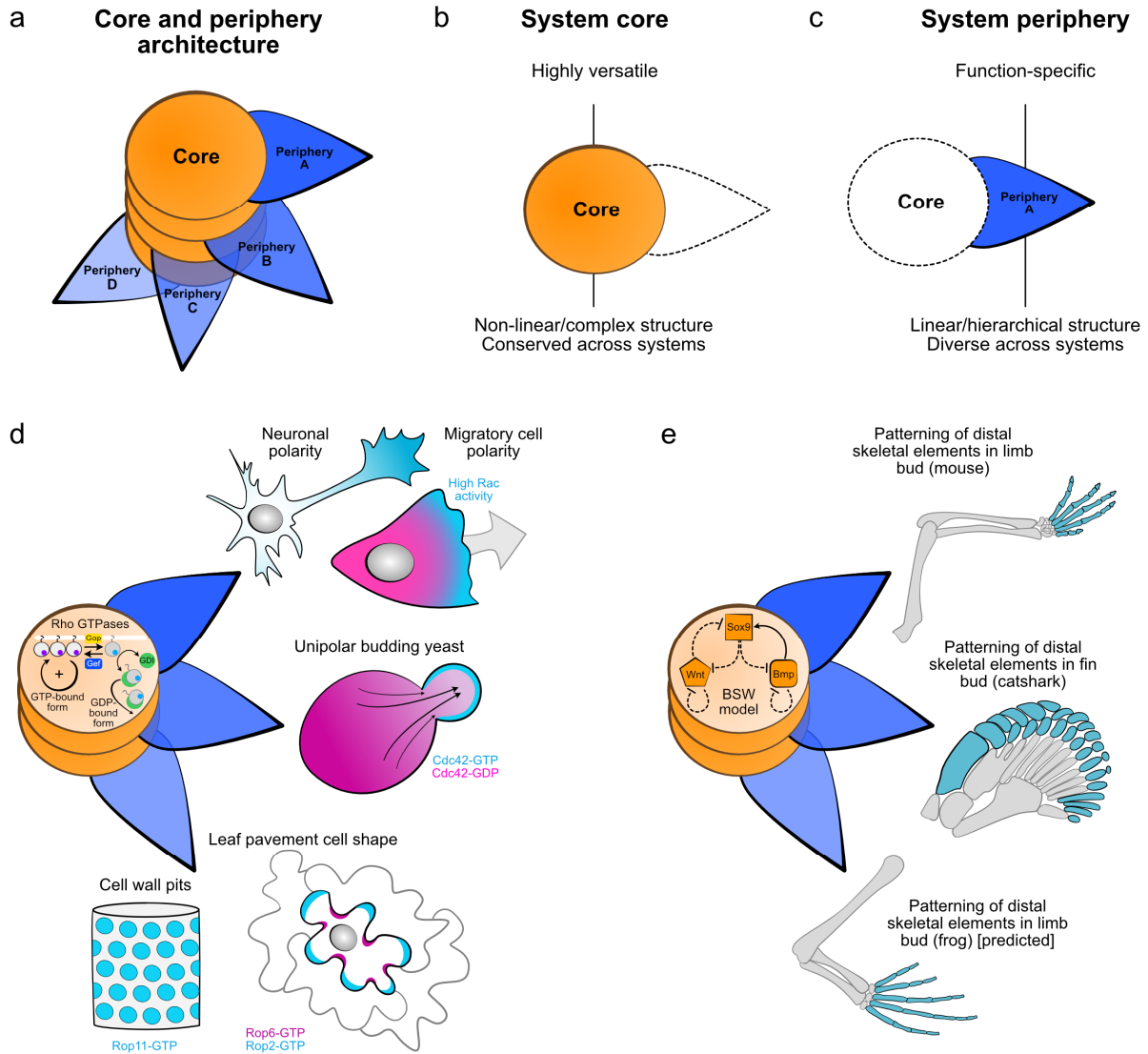

**Figure 2: The Core & Periphery (C&P) Hypothesis**

**(a-c)** Illustrations of the C&P system architecture. C&P systems are comprised of a versatile core (*orange disk*) and a function-specific periphery (*blue leaf*), which programs or specifies the core to produce one of its many possible behaviors. **(a)** The same core is reused (*stack of orange disks*) across many biological systems in combination with different peripheries (*blue leaves*) to implement different functions. **(b-c)** The differing roles of cores and peripheries in a C&P system imply differences in their structure and evolutionary conservation. **(d-e)** Illustrations of two independently evolved core implementations for the same core principle, namely Turing-style reaction-diffusion dynamics that can spontaneously generate spatio-temporal patterns (*46–49*). Each time an implementation is discovered in evolution, it can be reused repeatedly with variations in its periphery, which in the case of Turing systems might comprise expression levels, enzymatic activities, domain size, diffusion coefficients, external cues, and so on. This confers generality to both the principle and its implementation(s), as illustrated by: **(d)** Rho-family GTPase patterning systems employed in *Arabidopsis* metaxylem vessel cell wall pits (*146*) and leaf pavement cells (*147*), *Saccharomyces* budding (*148*), neuronal axon/dendrite polarity (*149, 150*), and leukocyte migration (*151*). **(e)** The Bmp-Sox9-Wnt (BSW) patterning system employed in mouse digit patterning (*134, 152*), catshark pectoral fin patterning (*153*), and frog limb patterning (figure adapted from (*153*)). In frogs, the BSW's Turing activity has not yet been shown, but the components are present (*154*).

Turing system) providing the "computation-like" behavior that underpins their versatility, whereas peripheries will tend to be structured hierarchically or linearly around their core. Furthermore, peripheries will often comprise multiple distinguishable subsets that perform different functions with respect to their core, such as specifying initial conditions, setting parameters or boundary conditions, or refining or canalizing the core's outputs.

As we will argue below, biological systems with a C&P architecture can readily diversify in evolution, since minor changes in the periphery enable efficient exploration of the large phenotypic landscape inherent in the core's versatility. This makes it possible for a core to remain conserved whilst being reused repeatedly across diverse biological systems, always in conjunction with a different periphery that has evolved to elicit a specific functionally useful output from the large behavioral repertoire of its core. The resulting widespread reuse of conserved cores makes them a suitable basis for general explanations of diverse biological phenomena. This is the generative principle at the heart of our proposal, and we will develop its characterization and implications more fully throughout this paper.

It should be emphasized that cores are not simply useful functional building blocks that can be recombined to create different systems, like modules or motifs (which are well-known and valuable concepts whose relation to C&P systems is further discussed in a later section). By contrast, a system core must possess the capacity to generate a variety of non-trivial behaviors *within itself*, similar to a computer (and as opposed to a basic electronic circuit motif). Cores are thus *reprogrammed* rather than merely *redeployed* across different systems.

General biological theories and explanations based on widespread conserved cores will commonly consist of two separable aspects. The first is the underlying **core principle**, a theoretical concept that explains how the core attains its inherent versatility. Core principles are best expressed and studied in conceptual, mathematical, or computational terms. In the case of Turing systems, the core principle is expressed through reaction-diffusion equations that show how, under certain circumstances, diffusion will amplify rather than smoothen out variations in the spatio-temporal distribution of interacting substances (*46–49*). The implications of this concept can be studied theoretically using mathematical and computational approaches, without direct reference to real biological systems (*51*, *54–57*).

The second and equally important part of a C&P-based explanation is the **core implementation**, which describes how a core principle is actually realized in biology. This is best accomplished in terms of the molecular or cellular components and mechanisms that make up the core and its peripheral regulators. Note that the same core principle may be implemented in multiple different ways if it is "discovered" independently in evolutionary history. Turing systems, for instance, have been implemented independently at the cellular scale based on Rho GTPases (Fig. 2d) and at the tissue scale based on morphogens, such as Bmp, Sox9 and Wnt (known as the BSW system) in the case of digit patterning (Fig. 2e) (*49*, *58*, *59*).

To be clear, however, our proposal is not based on the generality gained from parallel or convergent evolution of different implementations of the same principle, but rather on the potential of single core implementations to become widespread across biological systems and species, due to the large phenotypic space they provide for evolution to explore through modification of their peripheries. In cases where this potential has been realized, both the core principle *and* its implementation are relatively general. Unearthing this form of generality and illuminating how it might be harnessed to advance cell and developmental biology is our primary focus here, though we do also return to the evolutionary aspects of C&P architectures in more detail below.

In summary, we hypothesize that the C&P system architecture (1) is a general concept that can be applied to many qualitatively different biological systems across multiple scales, (2) entails a generative principle that explains how systems with such an architecture can spread widely and evolve great diversity, and (3) provides a template for theories that generalize across this diversity and are comprised of a core principle and its biological implementation. Together, these points constitute the **Core & Periphery hypothesis**, which we see as a suitable foundation for a novel conceptual framework of mesoscale biology.

### *C&P architectures are prevalent across mesoscale biological systems*

The question immediately arises whether cellular and developmental systems do in fact widely possess C&P architectures in nature, and indeed we found that it is possible and fruitful to frame several important biological phenomena from a C&P perspective. Rather than attempting to be comprehensive, we here focus on a number of examples across different biological scales that we consider pertinent to illustrate the framework's broad applicability and to further clarify key concepts. We also discuss the distinction of C&P from classical notions of modules or motifs at the end of this section. Note that, for the purpose of simplicity, we will henceforth often refer to *core implementations* simply as *cores*, but will continue to always refer to *core principles* specifically as such.

#### *Actomyosin and other cores of dynamic cellular organization*

The first example is the actomyosin cytoskeleton (Fig. 3a). It is well established that actin fibers and myosin motors together constitute a highly versatile platform employed by cells to perform a vast array of mechanical functions (*60–63*). The versatility of the actomyosin cytoskeleton makes it a core, and its various regulators and modulators (cross-linkers, nucleators, myosin phosphatases, etc.) compose the context-specific peripheries under which actomyosin will mediate functions such as cytokinesis, lamellipodia formation, or apical constriction. Note that cell geometry and the external mechanical forces acting on the cytoskeleton also form part of actomyosin's periphery, as they modulate the structure and dynamics of cytoskeletal assemblies even if they do not directly alter the biochemical properties or expression levels of actin or myosin molecules (*64–67*). Intriguingly,

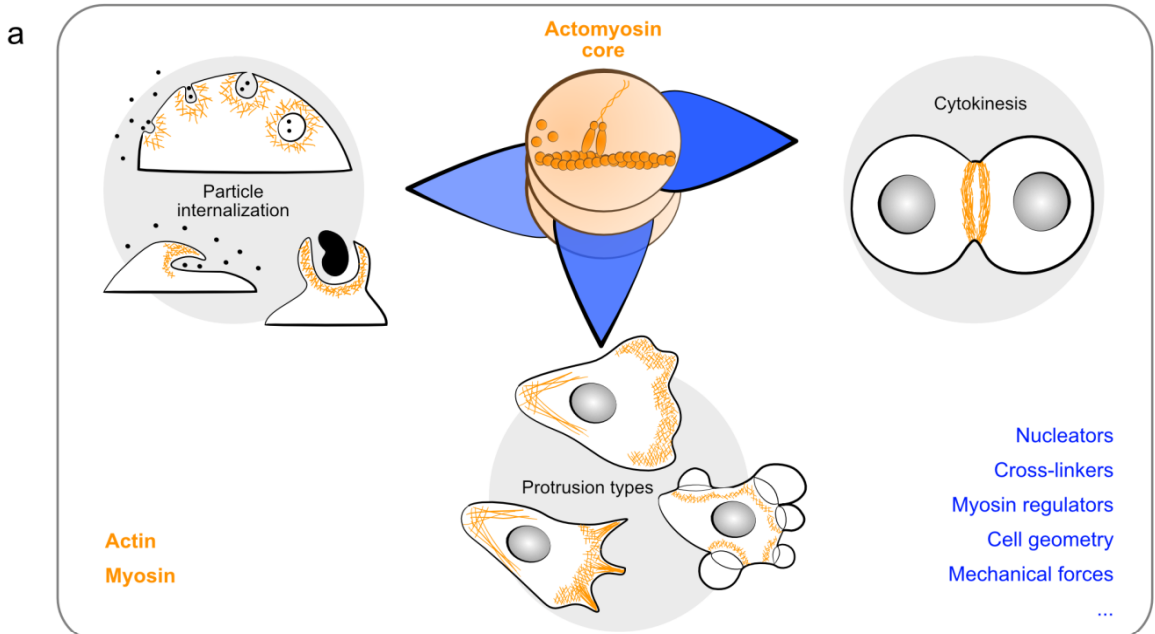

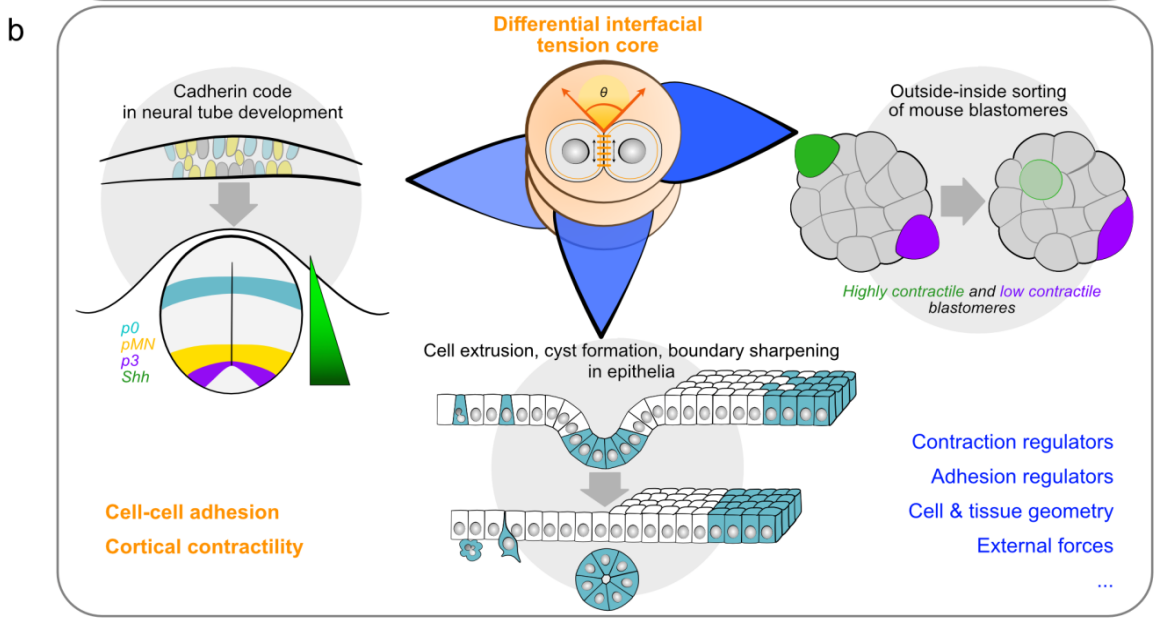

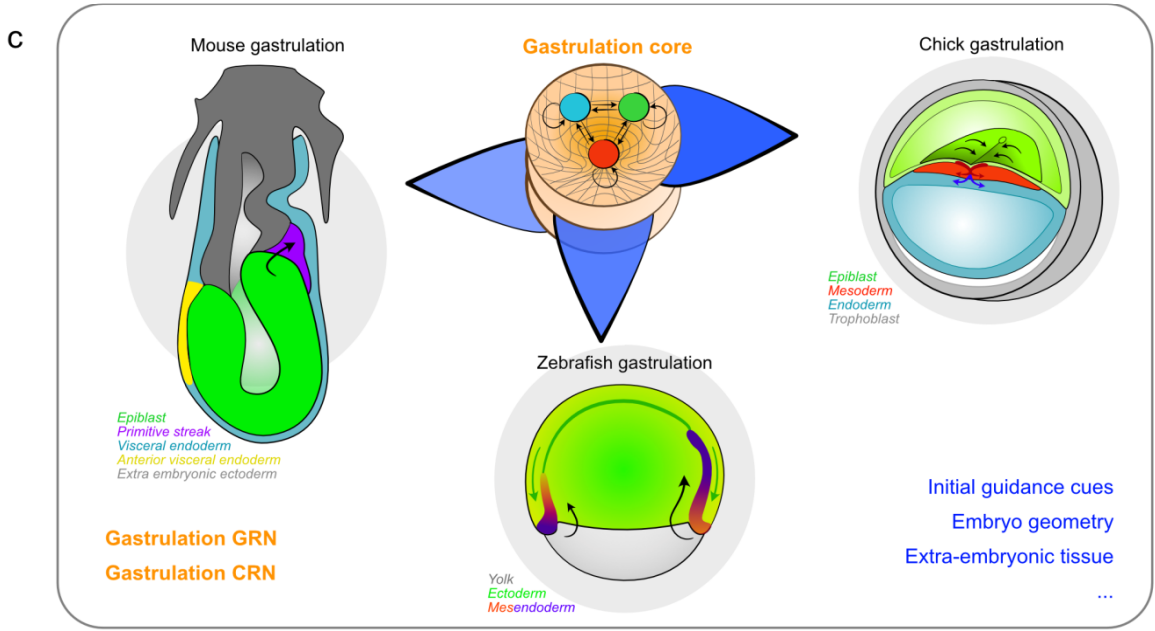

**Figure 3: Examples of C&P systems across multiple biological scales**

**(a)** Actomyosin as an example of a core at the (sub-)cellular scale. Actomyosin acts as a versatile platform for cellular mechanics and is reused in myriad ways by variation of its regulatory periphery (*60–62*). Illustrated are various forms of particle internalization, various types of cellular protrusions, and cytokinesis. **(b)** The differential interfacial tension core at the multi-cellular scale. The DITH and more recent extensions such as high Heterotypic Interfacial Tension (HIT) show how control over cellular contraction and adhesion results in a highly versatile toolkit for tissue morphogenesis (*81*, *84*, *155–157*). Depicted variations are the zebrafish neural tube adhesion code (adapted from (*158*)), cell sorting in mouse blastomeres (adapted from (*159*)), and lateral tension-based extrusion, cyst formation and boundary refinement in epithelia (*155*, *156*). **(c)** Gastrulation may employ a core at the embryonic scale, which is based on feedback between gene regulatory and cell regulatory networks (*101*). Illustrated are mouse, zebrafish and chick gastrulae, adapted from (*160*), (*161*) and (*162*), respectively.

though the multi-facetted versatility of actomyosin is well known, there is currently no general theory that explains the wide-ranging mechano-geometric capabilities of actomyosin-like active fiber systems. From a C&P perspective, the fact that actomyosin in its simplicity supports such a large variety of functions allows us to conjecture that such a theory – a core principle – should be obtainable. It is all the more exciting that recent work, namely the discovery of an elegant theory that universally predicts whether a disordered fiber-motor meshwork will contract or expand, shows that progress in this direction is possible (*68*) (see also (*69*)).

Similar to actomyosin, the tubulin-based cytoskeleton is also highly versatile and broadly utilized, including in the formation of the spindle, of cilia, and of cellular transport networks (*70–73*). It can therefore also be understood as a core, with its many regulators and in particular the microtubule-organizing centers forming the periphery. It will be interesting to explore if and how the core principles explaining the versatility of microtubules and of actomyosin overlap.

A third potential core on the cellular scale is the endomembrane system (*74–76*). It is increasingly appreciated that ER, Golgi and endolysosomal compartments are all highly diversified in their structure and function and can dynamically adjust their shapes, locations, interactions, and molecular compositions in myriad ways (*77–79*). It seems therefore plausible to view endomembrane compartments – alongside the key machinery mediating their segregation, localization and fusion – as a versatile core from which various functional organelles can be constructed or evolved via peripheral regulators.

Taken together, these examples illustrate that key cell-biological phenomena can be framed as C&P systems. We hope that taking such a perspective will support the development of new integrative theories that better generalize across the diverse ways in which cells organize themselves.

*Multi-cellular morphogenesis and the multi-scale nature of C&P*

Moving to the scale of cell collectives, there are a plethora of morphogenetic phenomena that demand explanation, ideally through general theories that account for many different morphogenetic behaviors at once and show how their diversity became accessible to evolution. For a broad class of spatial rearrangements within compact cell collectives, such a theory already exists. Its core principle goes back to D'Arcy Thompson's efforts to describe cell and tissue shapes using the physics of surface tension (*80*) and has since been greatly advanced, resulting in the Differential Interfacial Tension Hypothesis (DITH) (*81*). The DITH is a quantitative physical theory of how cell collectives undergo sorting, rearrangements, and shape changes by differentially modulating cell surface contractility and cell-cell adhesion (Fig. 3b) (*81–83*).

In animal development, this core principle is implemented by a combination of various adhesion proteins on the one hand and the cortical actomyosin meshwork on the other (*84, 85*). The periphery consists of signaling and gene regulatory networks that control adhesion protein expression and actomyosin contractility – and it again also includes geometric constraints and external forces. Notice that there is no single gene or protein that acts on its own to implement any aspect of the DITH core principle. Instead, both adhesion and contractility are mediated by the interactions of numerous proteins, each of which may also have other functions unrelated to interfacial tension. Thus, the DITH core implementation is not localized to individual genes or proteins and is better described in terms of higher-level components such as cell-cell junctions and the actomyosin cortex. As we will see below, it is often the case that implementations of cell- and tissue-scale C&P systems are best described in terms of motifs, structures, mechanisms, and processes, rather than by further reduction to the molecular level.

In this context, it is intriguing to note that one such higher-level component of the DITH core, namely the contractile cortex, is in fact mediated by a lower-level core, namely actomyosin. Put differently, a small subset of the many possible behaviors of the actomyosin core plays a key role in the implementation of the DITH core at the next higher level. It is therefore the actomyosin core *in conjunction with* a periphery that programs it to perform cortical contraction which together form a higher-level component of the DITH core implementation. Such multi-scale hierarchies are common, and although they can be challenging to grasp at first, they are a natural consequence of the applicability of C&P across multiple scales.

Two other potential cores at the multi-cellular scale bear mentioning. One is Epithelial-Mesenchymal Plasticity (EMP), often studied specifically in the context of the Epithelial-to-Mesenchymal Transition (EMT) (*86–88*). EMP is an abstraction over multiple correlated aspects of cell and tissue biology, including polarity (apico-basal vs. front-rear), transcriptional regulation (by E- or M-specific transcription factors), and cell mechanics (changes in adhesion, cell stiffness, and protrusion types) (*86, 89, 90*). EMP confers great morphogenetic versatility to cells and tissues by allowing them to dynamically shift in either direction along the E-M spectrum and to take on various intermediate states. However, while much is known about specific transitions in specific systems (especially EMT in cancer), a general theory of EMP biology remains to be discovered – a challenge that might benefit from a C&P-inspired approach.

The other tentative example is collective cell migration, which takes a multitude of different forms across development, cancer and wound healing, but always makes use of more or less the same components and lower-level cores (*91–93*). This lets us conjecture that general theories of collective cell migration phenomena may be attainable, although it is not clear if *all* such phenomena can be subsumed under a single core principle. Even so, an analysis of collectively moving cell populations as C&P systems – perhaps making use of concepts such as swarm intelligence (*94–97*) – could reveal interesting new biology.

The above examples show that one of the best-established theories of multi-cellular morphogenesis (the DITH) readily fits the C&P hypothesis, and that two other important morphogenetic phenomena (EMP and collective migration) might also be framed from this perspective as a basis for further investigation. Should such investigations prove successful, they may well lead to new theories that are more general and more predictive than the current patchwork of system-specific models.

*Gastrulation as an embryo-scale core*

The examples above already illustrated that cores can be implemented by multiple biological mechanisms that come together in a coordinated fashion to produce a higher-level system that possesses great versatility. This idea may extend even to the scale of entire embryos and major developmental events, such as vertebrate gastrulation (Fig. 3c) (*98, 99*). If the particulars are abstracted away, gastrulation can be viewed as a deeply ancestral chemo-mechanical partitioning system that couples cell fate decisions (mediated by Gene Regulatory Networks, GRNs) with cell rearrangements (mediated by Cell Regulatory Networks, CRNs) (*100, 101*), forming a gastrulation core whose versatility underlies the diversity of gastrulation processes observed across different species. Pre-patterned signals and embryonic geometry serve as initial and boundary conditions, which together with a host of genes that modulate various GRN and CRN parameters constitute the periphery.

Framing gastrulation as a C&P system may seem odd at first and extensive work will be required to substantiate this notion. However, two major experimental advances in the field already provide grounding empirical support for it. The first is the morphogenetic and transcriptional similarity of gastruloids generated from different species (*102*), which we will return to in a later section. The second comes from a recent result showing that chick gastrulation can be altered to mimic the gastrulation modes of reptiles, amphibians, or fish – simply by modulating the shape of the FGF-induced mesendoderm territory and/or the levels of EMT-driven ingression (*103*). Intriguingly, these findings can be explained with a single model that couples actomyosin activity to tissue flow (*104*), which hints at the existence of a simple but versatile core principle for gastrulation.

This example also highlights another important concept, which elsewhere has been described as **Developmental System Drift (DSD)** (*105*). So long as the relevant high-level behaviors (e.g. coordinated directional cell motion) are preserved, the specific low-level behaviors underlying them (e.g. the mode of cell migration) may change in evolution without disrupting a core implementation. Internalized mesodermal cells, for example, migrate away from the blastopore as a loose swarm in amniotes and fish, whereas they migrate as a coherent collective in frogs (*106*). In other words, substitutions in a core implementation can be silent with respect to the core principle.

*Comparison to modularity and combinatorial assembly*

Finally, it is worth distinguishing the C&P architecture from a well-established alternative way of decomposing complex biological systems, namely by splitting them into (structural) modules or network motifs. The central idea behind network motifs is that larger networks can be deconstructed into smaller units (e.g. feed-forward loops or double-negative-feedback loops) that perform a particular function (e.g. pulse generation or bistability) more or less independently from the rest of the network (*42*, *107*, *108*). Many biological networks show an over-abundance of a relatively small set of motifs compared to what would be expected by chance, indicating that these motifs serve as general building blocks from which larger networks are assembled (*42*, *107*). Similarly, cellular and developmental systems can be viewed as being composed of partially independent modules, such as deeply conserved signal transduction pathways (e.g. Notch signaling) which are redeployed and recombined time and again in evolution (*41*, *108–111*).

However, while such decompositions provide a path to generality at the module level (because modules are widely reused), they do not say much about how these building blocks tend to be organized into larger systems. For theories at the systems level, modularity thus provides diversity (by virtue of a combinatorial explosion), but it does not provide generality.

By contrast, the versatility of C&P systems is an emergent property of the core subsystem itself and does not arise merely from the combinatorial assembly of modular parts. Cores are thus chiefly distinct from modules/motifs in that they can perform *different* functions in different contexts – and in so doing make possible the discovery of common principles underpinning diverse biological phenomena. Generality is then found at the systems level, in the form of the core principle and the widespread occurrence of its implementation. To return to an analogy used earlier: the combinatorial assembly of modules and motifs resembles a circuit-based "engineering paradigm", whereas C&P resembles a computation-based "programming paradigm".

It must be stressed that these two ways of decomposing a biological system are not opposed or inconsistent. They are in fact complementary, as they can each provide their own form of simplicity and generality. Furthermore, C&P systems at one level can be combined into higher-level systems, including higher-level cores (cf. the role of cortical actomyosin in the DITH core). A particular core-periphery combination can thus effectively serve as a module. This complex relationship between combinatorial and emergent diversity remains to be explored, as do the relationships of C&P to various other concepts for which the term "module" has been employed (*112*).

### *An evolutionary explanation for the prevalence of C&P architectures*

Although we conceived the C&P hypothesis from a cell and developmental biology perspective, our exploration of the concept repeatedly led back to evolutionary considerations, as implied in a number of places above. Here, we discuss this aspect more explicitly.

At first glance, the generality posited by the C&P hypothesis may appear to be in tension with the randomness, historical contingency, and lack of foresight of evolution, which are sometimes thought to place a strict limit on the generality of biological theories (*15*, *113*). However, it is relatively straightforward to give a basic evolutionary account of C&P systems that resolves this apparent tension.

Consider the scenario in figure 4. Starting from a generic biological system that does not have a C&P architecture, classical evolution may at any point chance upon a new configuration that, by coincidence, does in fact have a rudimentary C&P structure. We term this event **core emergence**, and although it may be more or less rare (depending on the complexity of the core principle that is being "discovered"), there is nothing that systematically prevents its spontaneous occurrence. Indeed, self-organizing reaction-diffusion systems have been shown to emerge spontaneously in evolutionary simulations (*114*) and mono-functional patterning circuits can spontaneously gain multi-functionality (*115*).

Core emergence radically increases the evolutionary potential of the **core pioneer**, i.e. the organism that chanced upon the new core, as alterations in the periphery may now produce non-trivial new behaviors. By exploring and exploiting this newly unlocked phenotypic space, the descendants of the core pioneer have a greater chance of outcompeting other organisms in their own niche and of colonizing, invading, or constructing new niches. This mediates **core radiation** (spread of the new core implementation) coupled with **periphery individuation** (establishment of altered or novel peripheries that induce different core functions). Thus, the C&P hypothesis predicts that a core, once emerged, has the potential to spread widely and thereby confer generality to a theory describing it.

Put differently, cores spread because they provide a substantial increase in evolvability (*25*, *40*, *116*). As mentioned earlier, however, they do so not because they are self-contained modules that are readily *redeployed* (see our discussion on modules/motifs above, and refs (*32*, *42*)), but rather because they are dynamical subsystems that are readily *reprogrammed* to perform multiple different functions. Note also that core radiation does not require selection for evolvability itself (*116*, *117*), but only selection on the diverse evolved outputs of cores.

There will be many nuances and special cases associated with the evolutionary scheme proposed here. For instance, one might expect that cores will tend to lose their versatility due to drift or due to

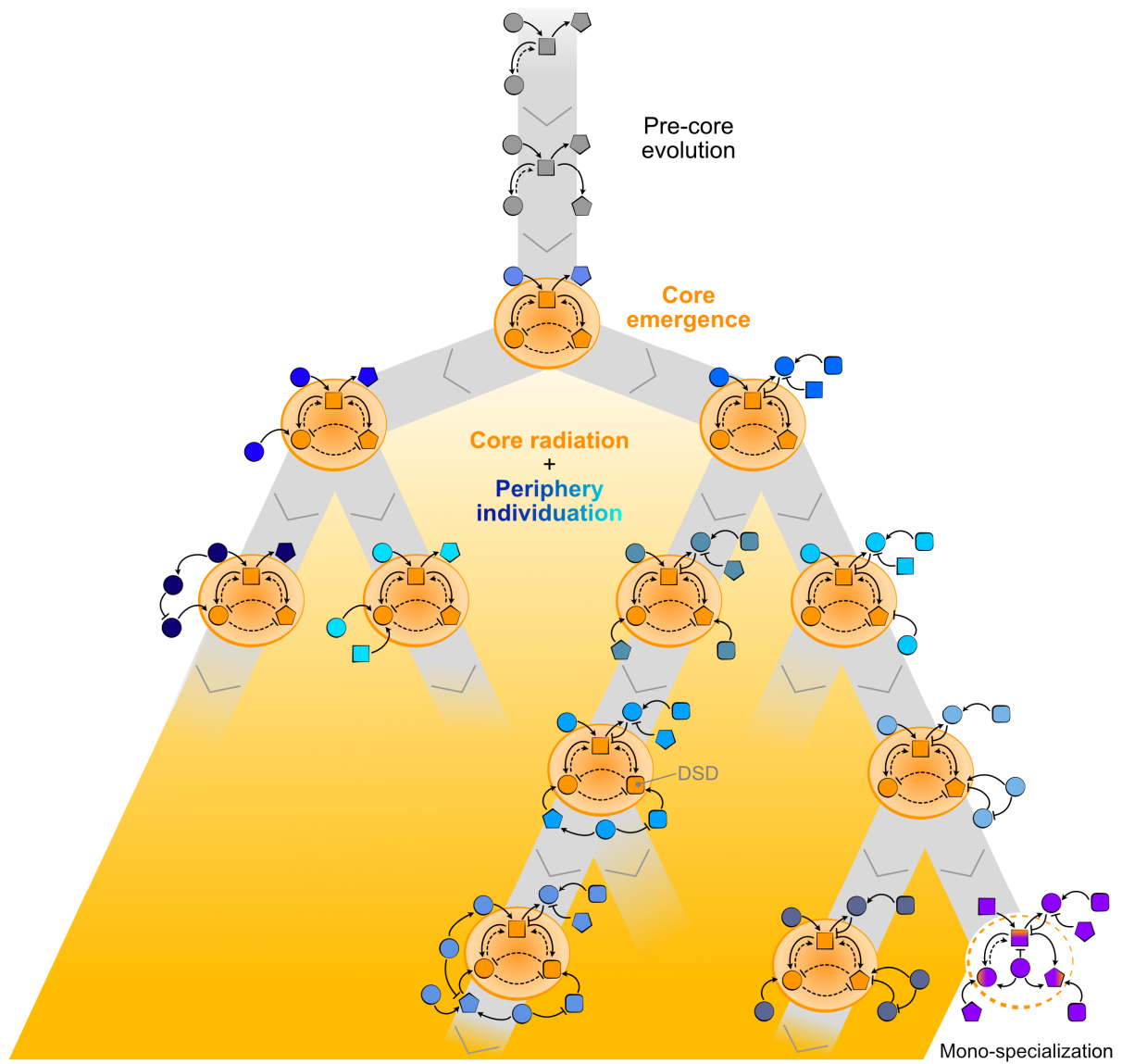

**Figure 4: The emergence and spread of C&P architectures in evolution**

In this illustration, biological systems are represented by components (*differently shaped nodes*) and their interactions (*black arrows*). Note that nodes could be genes or proteins, but could also be higher-level components, especially for systems above the subcellular scale. Ordinary evolution of a non-C&P system (*top, gray*) can lead to the spontaneous emergence of a new core, paired with its ancestral periphery. The increased evolutionary potential this confers to the core pioneer leads to core radiation (*conserved orange cores*) and periphery individuation (*changing blue peripheries*), both within a single evolving population and across speciation events. This accounts for the wide spread of cores and the relative generality of C&P-based theories. Note that developmental system drift (*DSD*) (*105*) can lead to the replacement of a core component with a functionally equivalent component without disrupting the core's ability to implement its core principle. Furthermore, under strong optimizing selection and especially in the presence of redundant copies of core components (e.g. after a genome duplication), changes in core components may become fixed at the expense of core versatility, leading to mono-specialization of the system and therefore loss of its C&P architecture (*bottom right*).

**mono-specialization**, wherein they trade off their versatility and "fuse" with a particular periphery in order to achieve greater optimization for one particular function (Fig. 4) (*114*). However, there are at least two factors that oppose these trends. First, the same core can be used to implement multiple different functions in the same organism, for example actomyosin being employed in both cell division and cell migration. Such multi-functionality is achieved through regulatory separation of different peripheries in space and/or time. Second, the generative versatility of cores can also implement *dynamical* versatility as a functional behavior in and of itself. For instance, the ability to dynamically switch between different modes of cell migration (e.g. between a mesenchymal and an amoeboid mode) enables a single cell to efficiently migrate in complex mechanical environments (102–104).

As soon as multi-functionality and dynamical versatility come into play, loss of core versatility can become detrimental to an organism. For instance, heavily optimizing one mode of migration at the cost of losing the ability to switch to the other mode may on balance decrease fitness. This may even favor a further increase in a core's versatility after its initial emergence, termed **core maturation**, as a core's role in multiple different functions creates pressure to shift any remaining function-specific aspects away from it and into the corresponding peripheries in order to reduce interference across functions. With these complexities in mind, which evolutionary mechanisms and dynamics ultimately explain the observed widespread entrenchment of extant cores becomes an open and interesting empirical question.

These scenarios only tackle one of multiple evolutionary dimensions of the C&P hypothesis. It would also be interesting to ask the converse question of how core emergence, radiation and maturation might bear on the dynamics of evolution itself. For example, core emergence is expected to be a rare event, but when it does occur it may enable comparably rapid evolutionary change as the newly gained versatility accelerates competition, niche invasion and niche construction (*118–123*). Core emergence may therefore be associated with punctuated equilibria, whereas gradual evolutionary change may primarily be driven by optimizations in the periphery (*119*, *124*, *125*).

Finally, we note that the continued integrity of cores across generations should make it possible to infer **core phylogenies**; reconstructions of the evolutionary history of a given core, including its radiation within and across derived species, and the diversification of its periphery. Recently, systematic strategies have been proposed for establishing homology between developmental processes (*126*) and between character identities (e.g. cell types) (*127–130*). Extending these ideas to C&P architectures may open up exciting new links between evolution and development.

*Some predictions and implications of the C&P hypothesis*

*General experimental predictions*

Conceptual frameworks do not on their own make specific predictions about particular biological systems (*18*). Nevertheless, some general predictions are implied by the C&P hypothesis that should apply to experiments on any biological system with a C&P architecture. This includes predictions about the different possible outcomes of perturbations (Fig. 5). Substantial perturbations of core components will invariably lead to a complete disruption of the system, resulting in a strong destructive phenotype that lacks organized biological behavior (Fig. 5a). Perturbations of the periphery (Fig. 5b) may yield very different outcomes depending on the specifics, ranging from no evident phenotype due to redundancy or canalization, to minor phenotypes resulting from small changes in system parameters or initial/boundary conditions, and even to major phenotypes in which the perturbed periphery fails entirely to elicit the proper behavior from the core. Such cases are not necessarily distinguishable from the catastrophic failure resulting from a perturbation in the core, though in some cases a distinction may be possible. Specifically, if the perturbation alters the periphery such that the core is shifted into a different functional regime, the resulting phenotype may resemble a phenomenon seen in some other biological system, as the core is effectively re-programmed to perform a (very crude) version of one of its other possible functions. This is the case in the experiments on chick gastrulation described above (*104*, *131*).

Intriguingly, another type of experiment becomes possible if the C&P structure of a system is sufficiently well understood: one can attempt to isolate or reconstitute a core without its periphery, keeping only the essential requirements for the core to function (Fig. 5c). This is difficult, but if successful one would expect to observe a **naive core behavior**, which should be invariant no matter the original biological source of the isolated core components. Based on the emerging concept of guided self-organization (*50*, *53*, *132*), we would expect such naive core behaviors to be highly unstable and sensitive to minor changes in experimental conditions, as is the case for example in Turing patterning systems that are not constrained by specific initial conditions or other external guidance cues (*56*, *133*, *134*). Artificial peripheries can subsequently be added back to a naive core in order to re-create existing (natural) core behaviors or to induce entirely new (synthetic) ones.

A well-established example of this is actomyosin, which if purified and placed in solution with ATP will produce dynamic, unstable, self-patterning meshworks (*69*, *135–137*). These can be guided to take specific forms using e.g. pre-patterned adhesion molecules on a surface (*138*, *139*). A less obvious yet very interesting case is found in gastrulation. After much trial and error, the right conditions have now been found for reconstituting at least part of the gastrulation core with a much-reduced periphery, resulting in a gastruloid (*140*, *141*). Remarkably, gastruloids turn out highly similar in terms of their morphogenetic and transcriptional dynamics regardless of the species from which they are derived – despite the vast differences in the natural gastrulae of these species (*102*).

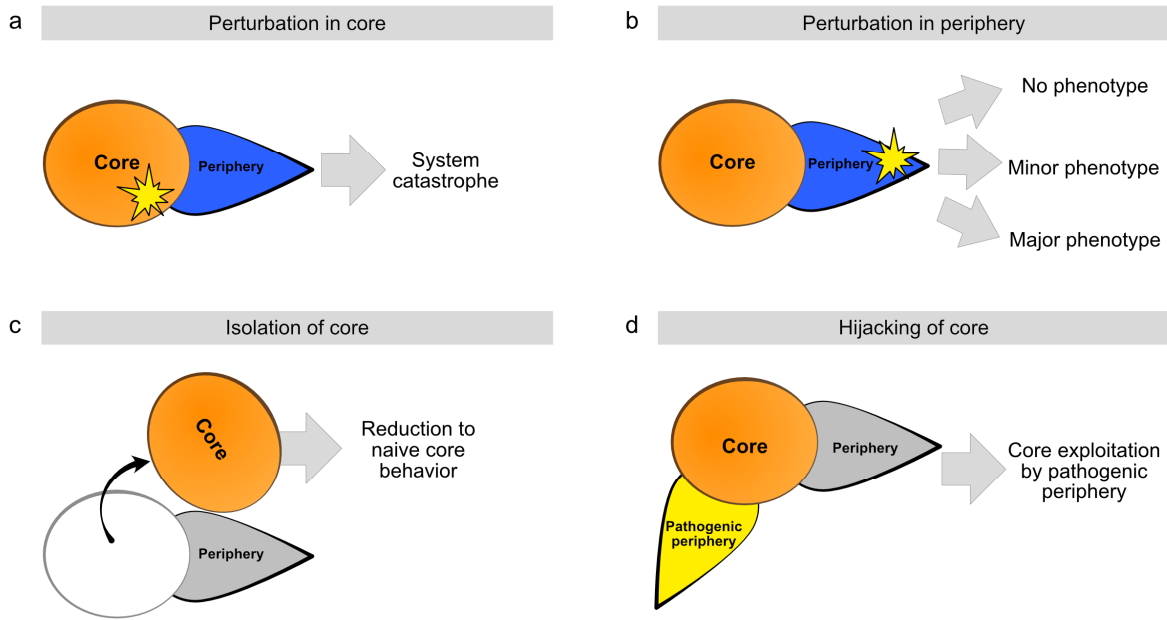

**Figure 5: General experimental predictions of the C&P hypothesis**

(a) Any substantial perturbation (yellow spark) of the core will disrupt the entire system, leading to a catastrophic failure and drastic phenotype. (b) The effects of substantial perturbations in the periphery can vary depending on the specifics, from no obvious phenotype if the system is robust to the perturbation, to minor phenotypes, and even to major phenotypes if the periphery is altered such that it fails to elicit the proper behavioral regime from the core. Such a major phenotype may not be distinguishable from the catastrophic failure resulting from a disruption in the core, so destructive phenotypes alone are not sufficient to conclude that a core component has been perturbed. (c) If the periphery can be removed or reduced – or equivalently if a core can be extracted or reconstituted (black arrow) – in such a way that only the basic permissive conditions for the core's functioning remain in place, the core is expected to display a naive behavior that is the same regardless of the original source from which the isolated core was obtained. (d) Viruses and cancers might be understood as pathogenic peripheries (or pathogenic alterations of existing peripheries) that reprogram host cores in ways that benefit them at the expense of their host.

This otherwise puzzling result is readily explicable under a C&P model of gastrulation (see Examples section), where gastruloid formation can be understood as the naive behavior of a common gastrulation core in a minimal periphery.

*Implications for the study of mesoscale biological systems*

An important role of conceptual frameworks is to guide practice in the field. Therefore, we here briefly outline a template for how to approach the study of complex mesoscale systems from a C&P perspective.

In a first step, the relevant core (or cores) should be identified. This is easiest if the core involved is already well-established in other systems, which will eventually become the norm given that each core underpins many different biological systems. If the presence of a hitherto unknown core is suspected, there are at least three strategies for identifying and characterizing it.

One strategy is to probe the structure and the dynamical properties of the system: cores are expected to have a complex and highly interconnected structure and to exhibit non-linear dynamics that effectively perform some kind of computation, whereas peripheries will be more linearly or hierarchically structured around their cores and exhibit simpler dynamics that combine, amplify, forward, or filter information.

A second strategy is to exploit the predicted responses of C&P systems to perturbations, discussed in the previous subsection (see Fig. 5a-c). Indeed, the ability to isolate a core and demonstrate its versatility by combining it with various synthetic peripheries may well become the experimental gold standard for core identification. Destructive phenotypes on the other hand must be interpreted with caution: while ablating a core component is always destructive, not all destructive phenotypes are due to ablation of a core component. This is because ablation of a peripheral component may be destructive as well, depending on how important it is in programming the core's behavior within the specific biological system under study.

The third strategy uses comparative approaches based on the evolutionary predictions of the C&P hypothesis. Sequence conservation may give a first hint as to which genes contribute to a core, since core components will usually be more highly conserved than peripheral components. However, such conservation may be very weak for cores at higher levels of organization, as they involve many genes that may each undergo some degree of Developmental System Drift (*105*). A far stronger source of evidence is the co-occurrence of all core components across different systems, as the absence of one core component would render the core non-functional. For higher-level cores, recent work on systematic ways of determining homology at the level of entire mechanisms rather than genes will be essential (*126*). One way or another, a proposed core must ultimately be shown to reoccur across several different biological systems and to perform different functions in systems with different peripheries. Intriguingly, such comparative studies of C&P systems can be done not only across different species, but also across different cell types or organs that utilize the same core for different purposes. This makes it possible to perform comparative studies within a single organism.

Once a core is identified, the next step is to understand how it works. Such understanding is of profound value because it generalizes across all systems that utilize the core in question. Classical experimental approaches can be applied to unravel the cellular or molecular underpinnings of the core implementation, whereas an interplay between quantitative experiments and theoretical modeling is likely key in revealing and understanding the core principle.

Finally, a complete understanding of a system of interest will also require an understanding of how the periphery programs the core. The most common ways in which a given core's behavior can be modulated by its various peripheries may generalize across all use-cases of the core, representing a sort of general "programming interface". Synthetic experiments and simulations wherein different peripheries are added to a core will be especially useful in elucidating these general aspects. However, peripheries in their entirety are by definition system-specific. Thus, a full understanding of any *particular* C&P system must ultimately encompass an idiosyncratic description of its periphery.

Coming back to cores, it should be noted that insights gained about them from studies on experimentally tractable model systems will also translate into systems that are ill-suited for foundational studies, such as many medically relevant systems. This simplifies the hard task of understanding such less tractable systems to the easier task of understanding how their particular peripheries are configuring an already well-characterized core. In other words, the C&P hypothesis facilitates systematic extrapolation from basic to applied science. In this context, we note that both cancer and viruses might be understood as **pathogenic peripheries** that exploit existing host cores by reprogramming them (Fig. 5d).

*Consolidating and building upon the C&P framework*

We have but scratched the surface of the ideas presented here, so extensive future work will be required to bring the promise of the C&P hypothesis to fruition.

One open conceptual question is whether individual components or mechanisms can always be clearly assigned to the core or to the periphery, the alternative being that **coreness** is a matter of degree, with some components or mechanisms contributing more to the core's versatility than others (and therefore also being more or less widely reused). It will be interesting to explore the distribution of such a coreness parameter across biological systems and to investigate how this relates to versatility and evolution.

A related goal is the construction of a mathematical toolset to effectively and quantitatively represent key aspects of C&P systems. This will give rise to metrics that can be data-mined across bioinformatics resources or measured in real and simulated systems to quantify the prevalence of C&P architectures in nature and to better understand their properties.

Another aspect of the research program suggested by the C&P hypothesis is the construction of an ontology (and eventually a phylogeny) of all major cores in nature. Indeed, if the ultimate aim of the gene-function paradigm is to discover the functions of every gene and the genes underlying every function, the ultimate aim of a C&P paradigm would be to discover all core principles and all of their major core implementations. The number of core principles is expected to be small compared to the number of biological systems in which they are employed, and whilst independent core implementations of principles that are easily evolved (such as Turing instabilities) may be numerous, the number of deeply ancestral and therefore truly widespread core implementations will likely be a small multiple of the number of core principles. One challenge in the construction of an ontology of cores will be the clear delineation of what is and what is not a core. It is therefore imperative to make the strategies for core identification discussed above (structure and dynamics, reductive and synthetic experiments, and evolutionary analysis) as concrete and quantitative as possible.

Finally, the C&P hypothesis is an integrative concept in the sense that it draws on and relates to many other theories and perspectives. We have already highlighted links to structural motifs and modules (*42*, *111*) and to evolutionary concepts such as evolvability (*25*, *40*, *117*). In addition, there are connections to guided self-organization (*50*, *53*), Dynamical Patterning Modules (DPMs) (*43*, *142*), dynamical modules more broadly (*112*, *143*), and Generative Entrenchment (*44*, *45*). Each of these concepts features some overlap and agreement with the C&P perspective, but also differs in its aims, scope, focus and implications. Furthermore, there are important aspects of mesoscale biology that C&P is not intended to address, most notably questions about cell types and tissue/organ identities, the ontology and evolution of which is better captured by frameworks such as kernels/plug-ins (*41*) and Character Identity Mechanisms (ChIMs) (*127*). Deeper analysis of the links between the various theories populating this emerging conceptual landscape will be a source of further advancements.

All of these goals will be challenging to achieve and pursuing them will no doubt reveal flaws and limitations that demand future revisions of the C&P framework as we have drafted it here. But even in its current early form, we believe that spending some time on an attempt to reframe one's favorite unsolved question or favorite model system from a C&P perspective can yield new insights and ideas.

## Conclusions

We sought to develop a conceptual framework to underpin general theories in the systems biology of cells, tissues and embryos. We propose that the Core & Periphery system architecture, wherein a versatile system core is functionally specified by the system periphery, can serve as the generative principle underpinning such a framework, since it implies that cores will tend to be widely reused and that theories of cores will therefore widely generalize. We have substantiated this hypothesis by discussing various examples, an evolutionary justification, and key strategies for identifying and characterizing cores. In our view, the complexity and diversity of mesoscale biological systems are the fundamental challenges that currently limit conceptual progress in cell and developmental biology. By separating the general from the idiosyncratic, the C&P hypothesis reveals hidden simplicity and generality, and thereby opens up a promising angle of attack on these central issues.

If successful, we expect that the C&P framework will have implications for applied biology as well. We already highlighted a potential link to medicine in the context of viruses and cancer. Another field that stands to benefit is synthetic biology (*132*). If a core is well understood, one should plausibly be able to exploit its versatility by programming it to perform a desired function through engineering of a custom periphery. In the long term, standardized versions of powerful cores and complementary peripheral toolsets could be "mass-produced" and serve as programmable platforms for C&P-driven biological engineering, not unlike how computers serve as platforms for software engineering. In addition, core principles are to some extent independent of their biological implementations, so it may also be possible to create "bio-mimetic cores" (i.e. mechanical or computational implementations of a core principle) whose versatility could drive advances in robotics and artificial intelligence.

The C&P framework is in many ways still in its infancy. Its theoretical basis will need to be refined, its implications elaborated, and its predictions made quantitative. Most importantly, these predictions will need to be empirically tested by experimental and computational means. We hope that the concepts, arguments and examples presented here will convince readers that such work is both interesting and important to pursue. More broadly, we hope that these ideas are taken to suggest that there remains much room for progress at the conceptual foundations of biology.

## Conflict of Interest Statement

The authors declare that they have no conflicts of interest with respect to this research.

## Author Contributions

JH and EG conceived the study. JH performed theoretical investigations with support from EG, SDR, JS, and RM. JH and EG wrote the manuscript. All authors reviewed and edited the manuscript.

## Acknowledgments

We are indebted to Luís Amaral (LA, Northwestern University), Darren Gilmour (DG, University of Zurich), Miki Ebisuya (ME, EMBL Barcelona), and Vikas Trivedi (VT, EMBL Barcelona) for their support. For discussion and feedback at various points during the conception and maturation of this work, we thank Andrew Kennard, Ben Helsen, Darren Gilmour, Elisabeth Kugler, Joergen Benjaminsen, Johannes Jaeger, Julie Theriot, Luís Amaral, Martin Gerlach, Marvin Albert, Matyas Bubna-Litic, Miki Ebisuya, Miquel Marin-Riera, Nade Abazova, Namid Stillman, Thomas Stoeger, and Xavier Diego. For feedback on an earlier version of the manuscript, we thank Adam Shellard, Andrew Kennard, James DiFrisco, Joergen Benjaminsen, and Namid Stillman.

JH and EG were partially supported by the University of Zürich (via DG). JH was also partially supported by the Northwestern Institute on Complex Systems (via LA), by the EMBL Scientific Visitor Programme (via JS, ME, and VT), by a Medical Research Council grant to RM (MRC, 558941), and by an EMBO long-term postdoctoral fellowship (ALTF 1284-2020). Work in RM's laboratory is supported by grants from the Medical Research Council (MR/S007792/1), Biotechnology and Biological Sciences Research Council (M008517, BB/T013044) and Wellcome Trust (102489/Z/13/Z).

# References

1. Y. Kashima, Y. Sakamoto, K. Kaneko, M. Seki, Y. Suzuki, A. Suzuki, Single-cell sequencing techniques from individual to multiomics analyses. *Exp Mol Med*. **52**, 1419–1427 (2020).

2. P. Pantazis, W. Supatto, Advances in whole-embryo imaging: a quantitative transition is underway. *Nat Rev Mol Cell Biol*. **15**, 327–339 (2014).

3. L. Przybyla, L. A. Gilbert, A new era in functional genomics screens. *Nat Rev Genet*. **23**, 89–103 (2022).

4. P. Villoutreix, What machine learning can do for developmental biology. *Development*. **148**, dev188474 (2021).

5. L. N. Waylen, H. T. Nim, L. G. Martelotto, M. Ramialison, From whole-mount to single-cell spatial assessment of gene expression in 3D. *Commun Biol*. **3**, 1–11 (2020).

6. C. D. Stern, Reflections on the past, present and future of developmental biology. *Developmental Biology*. **488**, 30–34 (2022).

7. M. Bizzarri, D. E. Brash, J. Briscoe, V. A. Grieneisen, C. D. Stern, M. Levin, A call for a better understanding of causation in cell biology. *Nat Rev Mol Cell Biol*. **20**, 261–262 (2019).

8. S. F. Gilbert, Developmental biology, the stem cell of biological disciplines. *PLOS Biology*. **15**, e2003691 (2017).

9. J. B. Wallingford, We Are All Developmental Biologists. *Developmental Cell*. **50**, 132–137 (2019).

10. C. D. Stern, The 'Omics Revolution: How an Obsession with Compiling Lists Is Threatening the Ancient Art of Experimental Design. *BioEssays*. **41**, 1900168 (2019).

11. B. A. Cohen, How should novelty be valued in science? *eLife*. **6**, e28699 (2017).

12. D. St Johnston, The Renaissance of Developmental Biology. *PLOS Biology*. **13**, e1002149 (2015).

13. L. Zon, Improving the visibility of developmental biology: time for induction and specification. *Development*. **146**, dev174631 (2019).

14. J. Jaeger, N. Monk, Everything flows. *EMBO reports*. **16**, 1064–1067 (2015).

15. D. C. Krakauer, J. P. Collins, D. Erwin, J. C. Flack, W. Fontana, M. D. Laubichler, S. J. Prohaska, G. B. West, P. F. Stadler, The challenges and scope of theoretical biology. *Journal of Theoretical Biology*. **276**, 269–276 (2011).

16. P. Nurse, Biology must generate ideas as well as data. *Nature*. **597**, 305–305 (2021).

17. T. S. Kuhn, *The structure of scientific revolutions* (Chicago University of Chicago Press, 1970), vol. 111.

18. I. Lakatos, *Falsification and the methodology of scientific research programmes* (Springer, 1976).

19. C. U. Moulines, The Nature and Structure of Scientific Theories. *Metatheoria* (2010) (available at http://ridaa.unq.edu.ar/handle/20.500.11807/2386).

20. S. M. Scheiner, Toward a Conceptual Framework for Biology. *The Quarterly Review of Biology*. **85**, 293–318 (2010).

21. W. E. Zamer, S. M. Scheiner, A Conceptual Framework for Organismal Biology: Linking Theories, Models, and Data. *Integrative and Comparative Biology*. **54**, 736–756 (2014).

22. S. Brenner, The genetics of Caenorhabditis elegans. *Genetics*. **77**, 71–94 (1974).

23. C. Nusslein-Volhard, E. Wieschaus, Mutations affecting segment number and polarity in Drosophila. *Nature*. **287**, 795–801 (1980).

24. U. Irion, C. Nüsslein-Volhard, Developmental genetics with model organisms. *Proceedings of the National Academy of Sciences*. **119**, e2122148119 (2022).

25. P. Alberch, From genes to phenotype: dynamical systems and evolvability. *Genetica*. **84**, 5–11 (1991).

26. K. Baverstock, The gene: An appraisal. *Progress in Biophysics and Molecular Biology*. **164**, 46–62 (2021).

27. K. Z. McKenna, R. Gawne, H. F. Nijhout, The genetic control paradigm in biology: What we say, and what we are entitled to mean. *Progress in Biophysics and Molecular Biology*. **169–170**, 89–93 (2022).

28. J. DiFrisco, J. Jaeger, Genetic Causation in Complex Regulatory Systems: An Integrative Dynamic Perspective. *BioEssays*. **42**, 1900226 (2020).

29. E. S. Welf, G. Danuser, Using Fluctuation Analysis to Establish Causal Relations between Cellular Events without Experimental Perturbation. *Biophysical Journal*. **107**, 2492–2498 (2014).

30. A. L. Tyler, F. W. Asselbergs, S. M. Williams, J. H. Moore, Shadows of complexity: what biological networks reveal about epistasis and pleiotropy. *BioEssays*. **31**, 220–227 (2009).

31. C. H. Chandler, S. Chari, I. Dworkin, Does your gene need a background check? How genetic background impacts the analysis of mutations, genes, and evolution. *Trends in Genetics*. **29**, 358–366 (2013).

32. G. P. Wagner, J. Zhang, The pleiotropic structure of the genotype–phenotype map: the evolvability of complex organisms. *Nat Rev Genet*. **12**, 204–213 (2011).

33. Y. Eguchi, G. Bilolikar, K. Geiler-Samerotte, Why and how to study genetic changes with context-dependent effects. *Current Opinion in Genetics & Development*. **58–59**, 95–102 (2019).

34. A. A. Hyman, Whither systems biology. *Philosophical Transactions of the Royal Society B: Biological Sciences*. **366**, 3635–3637 (2011).

35. M. W. Kirschner, The Meaning of Systems Biology. *Cell*. **121**, 503–504 (2005).

36. M. Bizzarri, A. Palombo, A. Cucina, Theoretical aspects of Systems Biology. *Progress in Biophysics and Molecular Biology*. **112**, 33–43 (2013).

37. R. Breitling, What is systems biology? *Frontiers in Physiology*. **1** (2010) (available at https://www.frontiersin.org/articles/10.3389/fphys.2010.00009).

38. L. Wolpert, Positional information and the spatial pattern of cellular differentiation. *J. Theor. Biol.* **25**, 1–47 (1969).

39. S. A. Newman, Generic physical mechanisms of tissue morphogenesis: A common basis for development and evolution. *J Evolution Biol*. **7**, 467–488 (1994).

40. M. Kirschner, J. Gerhart, Evolvability. *Proceedings of the National Academy of Sciences*. **95**, 8420–8427 (1998).

41. E. H. Davidson, D. H. Erwin, Gene Regulatory Networks and the Evolution of Animal Body Plans. *Science*. **311**, 796–800 (2006).

42. U. Alon, Network motifs: theory and experimental approaches. *Nat Rev Genet*. **8**, 450–461 (2007).

43. S. A. Newman, R. Bhat, Dynamical patterning modules: physico-genetic determinants of morphological development and evolution. *Phys. Biol.* **5**, 015008 (2008).

44. W. C. Wimsatt, "Chapter 7 - Robustness and Entrenchment - How the Contingent Becomes Necessary" in *Re-Engineering Philosophy for Limited Beings: Piecewise Approximations to Reality* (Harvard University Press, 2007), pp. 133–145.

45. W. C. Wimsatt, Generative entrenchment and the developmental systems approach to evolutionary processes. *Cycles of contingency: Developmental systems and evolution*, 219–237 (2001).

46. A. M. Turing, The chemical basis of morphogenesis. *Philosophical Transactions of the Royal Society of London. Series B, Biological Sciences*. **237**, 37–72 (1952).

47. A. Gierer, H. Meinhardt, A theory of biological pattern formation. *Kybernetik*. **12**, 30–39 (1972).

48. H. Meinhardt, A. Gierer, Pattern formation by local self-activation and lateral inhibition. *BioEssays*. **22**, 753–760 (2000).

49. S. Kondo, T. Miura, Reaction-Diffusion Model as a Framework for Understanding Biological Pattern Formation. *Science* (2010), doi:10.1126/science.1179047.

50. M. Prokopenko, Guided self-organization. *HFSP J.* **3**, 287 (2009).

51. A. L. Krause, M. A. Ellis, R. A. Van Gorder, Influence of Curvature, Growth, and Anisotropy on the Evolution of Turing Patterns on Growing Manifolds. *Bull Math Biol*. **81**, 759–799 (2019).

52. S. Werner, H. T.-K. Vu, J. C. Rink, Self-organization in development, regeneration and organoids. *Current Opinion in Cell Biology*. **44**, 102–109 (2017).

53. J. S. Morales, J. Raspopovic, L. Marcon, From embryos to embryoids: How external signals and self-organization drive embryonic development. *Stem Cell Reports*. **16**, 1039–1050 (2021).

54. L. Marcon, X. Diego, J. Sharpe, P. Müller, High-throughput mathematical analysis identifies Turing networks for patterning with equally diffusing signals. *eLife*. **5**, e14022 (2016).

55. X. Diego, L. Marcon, P. Müller, J. Sharpe, Key Features of Turing Systems are Determined Purely by Network Topology. *Phys. Rev. X*. **8**, 021071 (2018).

56. N. S. Scholes, D. Schnoerr, M. Isalan, M. P. H. Stumpf, A Comprehensive Network Atlas Reveals That Turing Patterns Are Common but Not Robust. *Cell Systems*. **9**, 243-257.e4 (2019).

57. R. A. Van Gorder, V. Klika, A. L. Krause, Turing conditions for pattern forming systems on evolving manifolds. *J. Math. Biol.* **82**, 4 (2021).

58. W. M. Bement, G. von Dassow, Single cell pattern formation and transient cytoskeletal arrays. *Current Opinion in Cell Biology*. **26**, 51–59 (2014).

59. L. Hubatsch, N. W. Goehring, "Chapter Eight - Intracellular morphogens: Specifying patterns at the subcellular scale" in *Current Topics in Developmental Biology*, S. Small, J. Briscoe, Eds. (Academic Press, 2020; https://www.sciencedirect.com/science/article/pii/S0070215319300936), vol. 137 of *Gradients and Tissue Patterning*, pp. 247–278.

60. R. Zaidel-Bar, G. Zhenhuan, C. Luxenburg, The contractome - a systems view of actomyosin contractility in non-muscle cells. *Journal of Cell Science*. **128**, 2209–2217 (2015).

61. P. Agarwal, R. Zaidel-Bar, Principles of Actomyosin Regulation In Vivo. *Trends in Cell Biology* (2018), doi:10.1016/j.tcb.2018.09.006.

62. C. Dekraker, E. Boucher, C. A. Mandato, Regulation and Assembly of Actomyosin Contractile Rings in Cytokinesis and Cell Repair. *The Anatomical Record*. **301**, 2051–2066 (2018).

63. K. B. Velle, L. K. Fritz-Laylin, Diversity and evolution of actin-dependent phenotypes. *Current Opinion in Genetics & Development*. **58–59**, 40–48 (2019).

64. V. D. Tran, S. Kumar, Transduction of cell and matrix geometric cues by the actin cytoskeleton. *Current Opinion in Cell Biology*. **68**, 64–71 (2021).

65. M. M. Kessels, B. Qualmann, Interplay between membrane curvature and the actin cytoskeleton. *Current Opinion in Cell Biology*. **68**, 10–19 (2021).

66. P. Lappalainen, T. Kotila, A. Jégou, G. Romet-Lemonne, Biochemical and mechanical regulation of actin dynamics. *Nat Rev Mol Cell Biol*, 1–17 (2022).

67. J. Mueller, G. Szep, M. Nemethova, I. de Vries, A. D. Lieber, C. Winkler, K. Kruse, J. V. Small, C. Schmeiser, K. Keren, R. Hauschild, M. Sixt, Load Adaptation of Lamellipodial Actin Networks. *Cell*. **171**, 188-200.e16 (2017).

68. J. M. Belmonte, M. Leptin, F. Nedelec, A theory that predicts behaviors of disordered cytoskeletal networks. *Mol Syst Biol*. **13**, 941 (2017).

69. G. H. Koenderink, E. K. Paluch, Architecture shapes contractility in actomyosin networks. *Curr Opin Cell Biol*. **50**, 79–85 (2018).

70. H. V. Goodson, E. M. Jonasson, Microtubules and Microtubule-Associated Proteins. *Cold Spring Harb Perspect Biol*. **10**, a022608 (2018).

71. S. Bodakuntla, A. S. Jijumon, C. Villablanca, C. Gonzalez-Billault, C. Janke, Microtubule-Associated Proteins: Structuring the Cytoskeleton. *Trends in Cell Biology*. **29**, 804–819 (2019).

72. K. Röper, Microtubules enter centre stage for morphogenesis. *Philosophical Transactions of the Royal Society B: Biological Sciences*. **375**, 20190557 (2020).

73. M. Dogterom, G. H. Koenderink, Actin–microtubule crosstalk in cell biology. *Nat Rev Mol Cell Biol*. **20**, 38–54 (2019).

74. D. S. Schwarz, M. D. Blower, The endoplasmic reticulum: structure, function and response to cellular signaling. *Cell. Mol. Life Sci.* **73**, 79–94 (2016).

75. J. Mathur, Review: Morphology, behaviour and interactions of organelles. *Plant Science*. **301**, 110662 (2020).

76. A.-C. Borchers, L. Langemeyer, C. Ungermann, Who's in control? Principles of Rab GTPase activation in endolysosomal membrane trafficking and beyond. *Journal of Cell Biology*. **220**, e202105120 (2021).

77. C. A. Lee, C. Blackstone, ER morphology and endo-lysosomal crosstalk: Functions and disease implications. *Biochimica et Biophysica Acta (BBA) - Molecular and Cell Biology of Lipids*. **1865**, 158544 (2020).

78. A. P. Liu, R. J. Botelho, C. N. Antonescu, The big and intricate dreams of little organelles: Embracing complexity in the study of membrane traffic. *Traffic*. **18**, 567–579 (2017).

79. C. Delevoye, M. S. Marks, G. Raposo, Lysosome-related organelles as functional adaptations of the endolysosomal system. *Current Opinion in Cell Biology*. **59**, 147–158 (2019).

80. D. W. Thompson, *On growth and form* (Cambridge university press Cambridge, 1942), vol. 2.

81. G. W. Brodland, The Differential Interfacial Tension Hypothesis (DITH): A Comprehensive Theory for the Self-Rearrangement of Embryonic Cells and Tissues. *Journal of Biomechanical Engineering*. **124**, 188–197 (2002).

82. G. W. Brodland, H. H. Chen, The Mechanics of Heterotypic Cell Aggregates: Insights From Computer Simulations. *Journal of Biomechanical Engineering*. **122**, 402–407 (2000).

83. G. W. Brodland, H. H. Chen, The mechanics of cell sorting and envelopment. *Journal of Biomechanics*. **33**, 845–851 (2000).

84. F. Fagotto, The cellular basis of tissue separation. *Development*. **141**, 3303–3318 (2014).

85. K. Röper, "Chapter Four - Integration of Cell–Cell Adhesion and Contractile Actomyosin Activity During Morphogenesis" in *Current Topics in Developmental Biology*, A. S. Yap, Ed. (Academic Press, 2015; https://www.sciencedirect.com/science/article/pii/S0070215314000180), vol. 112 of *Cellular Adhesion in Development and Disease*, pp. 103–127.

86. J. Yang, P. Antin, G. Berx, C. Blanpain, T. Brabletz, M. Bronner, K. Campbell, A. Cano, J. Casanova, G. Christofori, S. Dedhar, R. Derynck, H. L. Ford, J. Fuxe, A. García de Herreros, G. J. Goodall, A.-K. Hadjantonakis, R. Y. J. Huang, C. Kalcheim, R. Kalluri, Y. Kang, Y. Khew-Goodall, H. Levine, J. Liu, G. D. Longmore, S. A. Mani, J. Massagué, R. Mayor, D. McClay, K. E. Mostov, D. F. Newgreen, M. A. Nieto, A. Puisieux, R. Runyan, P. Savagner, B. Stanger, M. P. Stemmler, Y. Takahashi, M. Takeichi, E. Theveneau, J. P. Thiery, E. W. Thompson, R. A. Weinberg, E. D. Williams, J. Xing, B. P. Zhou, G. Sheng, Guidelines and definitions for research on epithelial–mesenchymal transition. *Nat Rev Mol Cell Biol*. **21**, 341–352 (2020).

87. H. Acloque, M. S. Adams, K. Fishwick, M. Bronner-Fraser, M. A. Nieto, Epithelial-mesenchymal transitions: the importance of changing cell state in development and disease. *J Clin Invest*. **119**, 1438–1449 (2009).

88. D. Pei, X. Shu, A. Gassama-Diagne, J. P. Thiery, Mesenchymal–epithelial transition in development and reprogramming. *Nat Cell Biol*. **21**, 44–53 (2019).

89. S. Tripathi, H. Levine, M. K. Jolly, The Physics of Cellular Decision Making During Epithelial–Mesenchymal Transition. *Annual Review of Biophysics*. **49** (2020), doi:10.1146/annurev-biophys-121219-081557.

90. C. P. Bracken, G. J. Goodall, The many regulators of epithelial–mesenchymal transition. *Nat Rev Mol Cell Biol*, 1–2 (2021).

91. P. Friedl, D. Gilmour, Collective cell migration in morphogenesis, regeneration and cancer. *Nat Rev Mol Cell Biol*. **10**, 445–457 (2009).

92. C. Norden, V. Lecaudey, Collective cell migration: general themes and new paradigms. *Current Opinion in Genetics & Development*. **57**, 54–60 (2019).

93. R. Alert, X. Trepat, Physical Models of Collective Cell Migration. *Annual Review of Condensed Matter Physics*. **11**, 77–101 (2020).

94. A. Shellard, R. Mayor, Rules of collective migration: from the wildebeest to the neural crest. *Philosophical Transactions of the Royal Society B: Biological Sciences*. **375**, 20190387 (2020).

95. A. Dowdell, P. Paschke, P. Thomason, L. Tweedy, R. H. Insall, Competition between chemoattractants causes unexpected complexity and can explain negative chemotaxis (2022), p. 2022.12.07.519354, , doi:10.1101/2022.12.07.519354.

96. M. L. Woods, C. Carmona-Fontaine, C. P. Barnes, I. D. Couzin, R. Mayor, K. M. Page, Directional Collective Cell Migration Emerges as a Property of Cell Interactions. *PLoS One*. **9** (2014), doi:10.1371/journal.pone.0104969.

97. I. D. Couzin, Collective cognition in animal groups. *Trends in Cognitive Sciences*. **13**, 36–43 (2009).

98. L. Solnica-Krezel, D. S. Sepich, Gastrulation: Making and Shaping Germ Layers. *Annual Review of Cell and Developmental Biology*. **28**, 687–717 (2012).

99. C. D. Stern, *Gastrulation: from cells to embryo* (CSHL Press, 2004).

100. N. Gorfinkiel, A. Martinez Arias, The cell in the age of the genomic revolution: Cell Regulatory Networks. *Cells & Development*. **168**, 203720 (2021).

101. B. Steventon, L. Busby, A. M. Arias, Establishment of the vertebrate body plan: Rethinking gastrulation through stem cell models of early embryogenesis. *Developmental Cell*. **56**, 2405–2418 (2021).

102. A. Kerim, T. Vikas, Studying evolution of the primary body axis in vivo and in vitro. *eLife*. **10** (2021), doi:10.7554/eLife.69066.

103. M. Chuai, G. Serrano Nájera, M. Serra, L. Mahadevan, C. J. Weijer, Reconstruction of distinct vertebrate gastrulation modes via modulation of key cell behaviors in the chick embryo. *Science Advances*. **9**, eabn5429 (2023).

104. M. Serra, G. S. Nájera, M. Chuai, V. Spandan, C. J. Weijer, L. Mahadevan, A mechanochemical model recapitulates distinct vertebrate gastrulation modes (2021), p. 2021.10.03.462928, , doi:10.1101/2021.10.03.462928.

105. J. R. True, E. S. Haag, Developmental system drift and flexibility in evolutionary trajectories. *Evolution & Development*. **3**, 109–119 (2001).

106. L. Solnica-Krezel, Conserved Patterns of Cell Movements during Vertebrate Gastrulation. *Current Biology*. **15**, R213–R228 (2005).

107. R. Milo, S. Shen-Orr, S. Itzkovitz, N. Kashtan, D. Chklovskii, U. Alon, Network Motifs: Simple Building Blocks of Complex Networks. *Science*. **298**, 824–827 (2002).

108. G. P. Wagner, M. Pavlicev, J. M. Cheverud, The road to modularity. *Nat Rev Genet*. **8**, 921–931 (2007).

109. V. Mireles, T. O. F. Conrad, Reusable building blocks in biological systems. *Journal of The Royal Society Interface*. **15**, 20180595 (2018).

110. G. Schlosser, Modularity and the units of evolution. *Theory Biosci.* **121**, 1–80 (2002).

111. L. H. Hartwell, J. J. Hopfield, S. Leibler, A. W. Murray, From molecular to modular cell biology. *Nature*. **402**, C47–C52 (1999).

112. J. Jaeger, N. Monk, Dynamical modules in metabolism, cell and developmental biology. *Interface Focus*. **11**, 20210011 (2021).

113. N. A. M. Monk, Elegant hypothesis and inelegant fact in developmental biology. *Endeavour*. **24**, 170–173 (2000).

114. I. Salazar-Ciudad, S. A. Newman, R. V. Sole, Phenotypic and dynamical transitions in model genetic networks I. Emergence of patterns and genotype-phenotype relationships. *Evol Dev*. **3**, 84–94 (2001).

115. A. Jiménez, J. Cotterell, A. Munteanu, J. Sharpe, A spectrum of modularity in multi-functional gene circuits. *Molecular Systems Biology*. **13**, 925 (2017).

116. R. A. Watson, "Evolvability" in *Evolutionary Developmental Biology: A Reference Guide*, L. Nuño de la Rosa, G. B. Müller, Eds. (Springer International Publishing, Cham, 2021; https://doi.org/10.1007/978-3-319-32979-6_184), pp. 133–148.

117. M. Pigliucci, Is evolvability evolvable? *Nat Rev Genet*. **9**, 75–82 (2008).

118. K. N. Laland, T. Uller, M. W. Feldman, K. Sterelny, G. B. Müller, A. Moczek, E. Jablonka, J. Odling-Smee, The extended evolutionary synthesis: its structure, assumptions and predictions. *Proceedings of the Royal Society B: Biological Sciences*. **282**, 20151019 (2015).

119. T. Peterson, G. B. Müller, Phenotypic Novelty in EvoDevo: The Distinction Between Continuous and Discontinuous Variation and Its Importance in Evolutionary Theory. *Evol Biol*. **43**, 314–335 (2016).

120. M. J. West-Eberhard, Developmental plasticity and the origin of species differences. *Proceedings of the National Academy of Sciences*. **102**, 6543–6549 (2005).

121. R. Lande, "Evolution of Phenotypic Plasticity in Colonizing Species" in *Invasion Genetics* (John Wiley & Sons, Ltd, 2016; https://onlinelibrary.wiley.com/doi/abs/10.1002/9781119072799.ch9), pp. 165–174.

122. C. L. Richards, O. Bossdorf, N. Z. Muth, J. Gurevitch, M. Pigliucci, Jack of all trades, master of some? On the role of phenotypic plasticity in plant invasions. *Ecology Letters*. **9**, 981–993 (2006).

123. P. Friedl, S. Alexander, Cancer Invasion and the Microenvironment: Plasticity and Reciprocity. *Cell*. **147**, 992–1009 (2011).

124. S. J. Gould, N. Eldredge, Punctuated equilibria: an alternative to phyletic gradualism. *Models in paleobiology*. **1972**, 82–115 (1972).

125. T. Pievani, How to Rethink Evolutionary Theory: A Plurality of Evolutionary Patterns. *Evol Biol*. **43**, 446–455 (2016).

126. J. DiFrisco, J. Jaeger, Homology of process: developmental dynamics in comparative biology. *Interface Focus*. **11**, 20210007 (2021).

127. J. DiFrisco, A. C. Love, G. P. Wagner, Character identity mechanisms: a conceptual model for comparative-mechanistic biology. *Biol Philos*. **35**, 44 (2020).

128. G. P. Wagner, The developmental genetics of homology. *Nat Rev Genet*. **8**, 473–479 (2007).

129. G. P. Wagner, *Homology, Genes, and Evolutionary Innovation* (Princeton University Press, 2014; https://www.degruyter.com/document/doi/10.1515/9781400851461/html).

130. D. Arendt, J. M. Musser, C. V. H. Baker, A. Bergman, C. Cepko, D. H. Erwin, M. Pavlicev, G. Schlosser, S. Widder, M. D. Laubichler, G. P. Wagner, The origin and evolution of cell types. *Nat Rev Genet*. **17**, 744–757 (2016).

131. M. Chuai, G. S. Nájera, M. Serra, L. Mahadevan, C. J. Weijer, Reconstruction of distinct vertebrate gastrulation modes via modulation of key cell behaviours in the chick embryo (2021), p. 2021.10.03.462938, , doi:10.1101/2021.10.03.462938.

132. J. Hartmann, R. Mayor, Self-organized collective cell behaviors as design principles for synthetic developmental biology. *Seminars in Cell & Developmental Biology* (2022), doi:10.1016/j.semcdb.2022.04.009.

133. P. K. Maini, T. E. Woolley, R. E. Baker, E. A. Gaffney, S. S. Lee, Turing's model for biological pattern formation and the robustness problem. *Interface Focus*. **2**, 487–496 (2012).

134. J. Raspopovic, L. Marcon, L. Russo, J. Sharpe, Digit patterning is controlled by a Bmp-Sox9-Wnt Turing network modulated by morphogen gradients. *Science*. **345**, 566–570 (2014).

135. Sonal, K. A. Ganzinger, S. K. Vogel, J. Mücksch, P. Blumhardt, P. Schwille, Myosin-II activity generates a dynamic steady state with continuous actin turnover in a minimal actin cortex. *Journal of Cell Science*. **132**, jcs219899 (2018).

136. R. D. Mullins, S. D. Hansen, In vitro studies of actin filament and network dynamics. *Current Opinion in Cell Biology*. **25**, 6–13 (2013).

137. M. Murrell, P. W. Oakes, M. Lenz, M. L. Gardel, Forcing cells into shape: the mechanics of actomyosin contractility. *Nat Rev Mol Cell Biol*. **16**, 486–98 (2015).

138. A.-C. Reymann, J.-L. Martiel, T. Cambier, L. Blanchoin, R. Boujemaa-Paterski, M. Théry, Nucleation geometry governs ordered actin networks structures. *Nature Mater*. **9**, 827–832 (2010).

139. A. C. Reymann, R. Boujemaa-Paterski, J. L. Martiel, C. Guerin, W. Cao, H. F. Chin, E. M. De La Cruz, M. Thery, L. Blanchoin, Actin network architecture can determine myosin motor activity. *Science*. **336**, 1310–4 (2012).

140. A. M. Arias, Y. Marikawa, N. Moris, Gastruloids: Pluripotent stem cell models of mammalian gastrulation and embryo engineering. *Developmental Biology*. **488**, 35–46 (2022).

141. S. C. van den Brink, A. van Oudenaarden, 3D gastruloids: a novel frontier in stem cell-based in vitro modeling of mammalian gastrulation. *Trends in Cell Biology*. **31**, 747–759 (2021).

142. S. A. Newman, R. Bhat, Dynamical patterning modules: a "pattern language" for development and evolution of multicellular form. *Int. J. Dev. Biol*. **53**, 693–705 (2009).

143. J. Jaeger, N. Monk, Dynamical Modularity of the Genotype-Phenotype Map (2019), , doi:10.31219/osf.io/vfz4t.

144. J. DiFrisco, Toward a theory of homology: development and the de-coupling of morphological and molecular evolution. *The British Journal for the Philosophy of Science* (2021), doi:10.1086/714959.

145. N. Shubin, C. Tabin, S. Carroll, Deep homology and the origins of evolutionary novelty. *Nature*. **457**, 818–823 (2009).

146. Y. Nagashima, S. Tsugawa, A. Mochizuki, T. Sasaki, H. Fukuda, Y. Oda, A Rho-based reaction-diffusion system governs cell wall patterning in metaxylem vessels. *Sci Rep*. **8**, 11542 (2018).

147. Y. Fu, Y. Gu, Z. Zheng, G. Wasteneys, Z. Yang, Arabidopsis Interdigitating Cell Growth Requires Two Antagonistic Pathways with Opposing Action on Cell Morphogenesis. *Cell*. **120**, 687–700 (2005).

148. A. B. Goryachev, A. V. Pokhilko, Dynamics of Cdc42 network embodies a Turing-type mechanism of yeast cell polarity. *FEBS Letters*. **582**, 1437–1443 (2008).

149. M. Fivaz, S. Bandara, T. Inoue, T. Meyer, Robust Neuronal Symmetry Breaking by Ras-Triggered Local Positive Feedback. *Current Biology*. **18**, 44–50 (2008).

150. J. Ng, T. Nardine, M. Harms, J. Tzu, A. Goldstein, Y. Sun, G. Dietzl, B. J. Dickson, L. Luo, Rac GTPases control axon growth, guidance and branching. *Nature*. **416**, 442–447 (2002).

151. O. D. Weiner, P. O. Neilsen, G. D. Prestwich, M. W. Kirschner, L. C. Cantley, H. R. Bourne, A PtdInsP3- and Rho GTPase-mediated positive feedback loop regulates neutrophil polarity. *Nat Cell Biol*. **4**, 509–513 (2002).

152. R. Sheth, L. Marcon, M. F. Bastida, M. Junco, L. Quintana, R. Dahn, M. Kmita, J. Sharpe, M. A. Ros, Hox Genes Regulate Digit Patterning by Controlling the Wavelength of a Turing-Type Mechanism. *Science*. **338**, 1476–1480 (2012).

153. K. Onimaru, L. Marcon, M. Musy, M. Tanaka, J. Sharpe, The fin-to-limb transition as the re-organization of a Turing pattern. *Nat Commun*. **7**, 11582 (2016).

154. A. Satoh, T. Endo, M. Abe, N. Yakushiji, S. Ohgo, K. Tamura, H. Ide, Characterization of Xenopus digits and regenerated limbs of the froglet. *Developmental Dynamics*. **235**, 3316–3326 (2006).

155. L. Canty, E. Zarour, L. Kashkooli, P. François, F. Fagotto, Sorting at embryonic boundaries requires high heterotypic interfacial tension. *Nat Commun*. **8**, 157 (2017).

156. C. Bielmeier, S. Alt, V. Weichselberger, M. La Fortezza, H. Harz, F. Jülicher, G. Salbreux, A.-K. Classen, Interface Contractility between Differently Fated Cells Drives Cell Elimination and Cyst Formation. *Current Biology*. **26**, 563–574 (2016).

157. N. C. Heer, A. C. Martin, Tension, contraction and tissue morphogenesis. *Development*. **144**, 4249–4260 (2017).

158. T. Y.-C. Tsai, M. Sikora, P. Xia, T. Colak-Champollion, H. Knaut, C.-P. Heisenberg, S. G. Megason, An adhesion code ensures robust pattern formation during tissue morphogenesis. *Science* (2020), doi:10.1126/science.aba6637.

159. J.-L. Maître, H. Turlier, R. Illukkumbura, B. Eismann, R. Niwayama, F. Nédélec, T. Hiiragi, Asymmetric division of contractile domains couples cell positioning and fate specification. *Nature*. **536**, 344–348 (2016).

160. R. M. Arkell, P. P. L. Tam, Initiating head development in mouse embryos: integrating signalling and transcriptional activity. *Open Biology*. **2**, 120030.

161. D. Pinheiro, C.-P. Heisenberg, "Chapter Twelve - Zebrafish gastrulation: Putting fate in motion" in *Current Topics in Developmental Biology*, L. Solnica-Krezel, Ed. (Academic Press, 2020; https://www.sciencedirect.com/science/article/pii/S0070215319300845), vol. 136 of *Gastrulation: From Embryonic Pattern to Form*, pp. 343–375.

162. J. G. Wittig, A. Münsterberg, The Early Stages of Heart Development: Insights from Chicken Embryos. *Journal of Cardiovascular Development and Disease*. **3**, 12 (2016).

# Glossary of Terms

| | |
|---|---|
| **conceptual framework** | A set of interconnected high-level concepts that span across a research field and serve as a foundation for constructing and reasoning about explanations and theories, as well as for designing and interpreting experiments to support or test such explanations. |
| **gene-function paradigm** | A **conceptual framework** for cell and developmental biology in which biological functions are ascribed to and explained by genes. Research under this framework is focused on the construction of gene-function maps (often based on mutant-phenotype relationships) and the elucidation of molecular mechanisms that implement particular gene functions. |
| **general / generality** | General biological explanations or theories are those that to some extent apply across different biological (and optionally non-biological) systems, as opposed to idiosyncratic explanations that are highly context-specific. Both types of explanation have their virtues: generalization usually reduces precision but enables understanding and prediction across a much wider range of contexts. Note that we use the term "general" not in the sense of "universal" (which we would take to mean generality across *all* biological systems), but rather in the sense of "relatively general", i.e. generality within a certain class of systems or range of contexts. |
| **generative principle** | A principle that explains how a set of facts, things or expressions were (or can be) generated. Powerful generative principles are relatively simple and yet describe the generation of large sets, often through repeated application of some process, as is the case in evolutionary theory. Note that the **gene-function paradigm** does not constitute a generative principle that is powerful in this sense, as it views functional diversity to be encoded in *equivalent* genetic diversity rather than being *generated* dynamically (*26–28, 105, 126, 144, 145*). |
| **versatility** | Here used to denote the capacity to perform multiple different functions. Thus, a biological (sub-)system that is highly versatile is one that can perform a wide range of different functions. Both quantitative and qualitative versatility (functions that differ in magnitude and functions that differ in kind, respectively) are relevant, but *qualitatively* different functions are of much greater import in our argument. Our use of the term must be distinguished from the ability to perform the same function in many different contexts. Thus, a CPU is versatile under this definition (because it can be *reprogrammed* to perform different functions), whereas a single electrical element such as a resistor is not (even though it can be *redeployed* in the construction of many different circuits). |
| **Core & Periphery (C&P) system / architecture** | A system (or system architecture) that features an intrinsically **versatile system core** and a complementary **system periphery** that programs the core to perform a specific function. |
| **system core** | A subset of a biological system that has the intrinsic capacity to generate a wide range of non-trivially different behaviors. As a consequence, cores tend to be frequently reused to implement different functions across various biological systems. |
| **system periphery** | The subset of a biological system that is complementary to the **system core** and can be thought to trigger or program the core such that it performs one specific behavior out of the many in its **versatile** repertoire. |
| **core principle** | A conceptual, mathematical or computational explanation of how the **versatility** of a **system core** emerges. Core principles can be abstract; they need not make reference to a particular biological system and can be realized independently by different **core implementations**. |
| **core implementation** | A particular biological realization of a **core principle**, described in terms of actual biological components, mechanisms and processes. A core principle can have multiple (independently evolved) implementations, each of which comes with great evolutionary potential and will therefore be reused multiple times across biology. |
| **Core & Periphery (C&P) hypothesis** | The hypothesis that **C&P systems** are prevalent at the biological mesoscale due to the evolutionary potential imparted by the inherent **versatility** of the **system core**, which is programmed by diverse **system peripheries** to perform different functions. This in turn implies that the study of cores can lead to **general** theories comprised of a **core principle** and its **core implementation(s)**, making the C&P hypothesis an attractive candidate for a novel **conceptual framework** of mesoscale biology. |
| **Developmental System Drift (DSD)** | A process wherein the molecular or mechanistic underpinnings of a developmental outcome diverge in evolution whilst the outcome itself remains relatively conserved (*105*). In C&P systems, especially in those above the cellular scale, DSD can occur both in the core implementation and in the periphery. |
| **core emergence** | A serendipitous evolutionary event that transforms a biological system which does not possess a C&P architecture into one that does. This spontaneous emergence of a core occurs in a **core pioneer** and endows it with increased evolutionary potential. |
| **core pioneer** | The organism within which a **core emergence** event takes place. The resulting increase in evolutionary potential confers to its descendants a higher chance of outcompeting other organisms in their own niche and of invading or constructing new niches, due to the large phenotypic space that is now readily accessible through modifications in the periphery of the new core. |
| **core radiation** | The spread of a core following **core emergence**, driven by the evolutionary success of the **core pioneer**'s descendants. Cores radiate both through being co-opted by different peripheries to perform different functions in different descendent species and through being employed repeatedly within a single descendent species through temporal or spatial regulation of the periphery. |
| **periphery individuation** | The complementary process to **core radiation**. While the core remains largely conserved as it radiates, different peripheries evolve through modification of the initial periphery (or occasionally through co-option by an entirely different periphery) to exploit the different functional uses provided by the core's versatility. |
| **mono-specialization** | An evolutionary process wherein fitness is increased by optimizing one particular function of a core at the cost of its versatility. If this occurs, the core and its periphery "fuse" and the system loses its C&P structure. Multi-functionality of a core within the same organism and selection for dynamical versatility can act to prevent mono-specialization. |
| **core maturation** | The process whereby a rudimentary core further increases in versatility because it is reused multiple times for different functions within the same organism, which creates a selection pressure to further separate the versatility-conferring aspects from the specificity-imposing aspects of the system into core and periphery, respectively. |
| **core phylogeny** | A reconstruction of the evolutionary history of a core implementation, from its multiple extant uses back to its universal common ancestor, i.e. the ancestral core that emerged in the **core pioneer**. The defining feature that remains consistent across a core's evolutionary history is its ability to implement the relevant **core principle**. In practice, core phylogenies may be traced based on the core's components, architecture, and dynamical properties, see (*126*). |
| **naive core behavior** | The behavior exhibited by a core when it is isolated into an environment with only a minimal (permissive) periphery. Naive core behaviors are invariant to the original biological systems from which the core was isolated. They will usually be unstable and sensitive to their context – and can thus readily be reprogrammed by the addition of (synthetic) peripheries. Note that the naive behavior of a core need not be related to the ancestral function performed by the core when it first emerged in evolution, as **core emergence** more plausibly occurs through the split of a system into core and periphery rather than the *de novo* generation of a core in the absence of any periphery (see section on evolution). |
| **pathogenic periphery** | A periphery that reprograms a core encoded by some host organism in such a way as to benefit a pathogen. Both viruses and cancers may act as pathogenic peripheries. Note that the generality/conservation of core implementations may facilitate cross-species transmission. |
| **coreness** | A parameter expressing how much a component or mechanism that is part of a C&P system contributes to the core's versatility. Coreness may be binary or bimodal in nature, enabling a clear delineation between core and periphery. However, it may also be more continuous, with some components or mechanisms not being essential to the core but still providing additional versatility when present, and therefore still being more widely reused and more conserved than other, entirely peripheral components. |